\documentclass[twocolumn,twoside,slac_two]{revtex4}
\usepackage{graphicx}
\usepackage{fancyhdr}
\begin{document}

\title{Background fitting of Fermi GBM observations}

%

\author{D. Sz\'{e}csi$^{1,2,3}$, Z. Bagoly$^{1,4}$, J. K\'{o}bori$^{1}$, L.G. Bal\'{a}zs$^{1,2}$, I. Horv\'{a}th$^{4}$}
\affiliation{$^{1}$E\"{o}tv\"{o}s University, Budapest, Hungary}
\affiliation{$^{2}$MTA CSKF Konkoly Observatory, Budapest, Hungary}
\affiliation{$^{3}$Argelander Institut f\"{u}r Astronomy, Bonn, Germany}
\affiliation{$^{4}$Bolyai Military University, Budapest, Hungary}
%

\begin{abstract}
The Fermi Gamma-ray Burst Monitor (GBM) detects gamma-rays in the energy range
8 keV - 40 MeV.  We developed a new background fitting process of these
data, based on the motion of the satellite.  Here we summarize 
this method, called \textsl{Direction Dependent Background Fitting} (DDBF),
regarding the GBM triggered catalog. We also give some preliminary results 
and compare the duration parameters with the 2-years Fermi Burst Catalog.
	
\end{abstract}

\maketitle

\thispagestyle{fancy}


\section{Introduction}

	Fermi has specific proper motion when surveying the sky. It is designed to catch
	gamma-ray bursts in an effective way.  However, bursts can have a varying background, especially in the case of the
	Autonomous Repoint Request (ARR).
	Modeling this with a polynomial function of time is not efficient in many cases.  Here we present the
	effect of these special moving feature, and we
	define direction dependent underlying variables. We use them 
	to fit a general multidimensional linear function for the background.
	
	Note that here we give a short summary of the method and some preliminary results.
	The detailed description of the Direction Dependent Background Fitting (DDBF) algorithm and
	the final results will be published soon \citep{AA}.

\section{Fermi lightcurves with varying background}

	Fermi's slewing algorithm is quite complex. It is designed to optimize the observation of
	the Gamma-Ray Sky. In Sky Survey Mode, the satellite rocks around the
	zenith within $\pm 50^{\circ}$, and the pointing alternates between the
	northern and southern hemispheres each orbit
	\citep{Meegan,Fitzpatrick}. 
	In Autonomous Repoint Request (ARR) mode, it turns toward the burst and stays there
	for hours.	
	12 NaI detectors are placed such a way
	that the entire hemisphere is observable with them at the same time.

	The GBM data, which we use in our analysis (called CTIME), are
	available at 8 energy channels, with $0.064$-second for triggered and $0.264$-second
	resolution for non-triggered mode. The
	position data is available in $30$-second resolution.  This data were
	evenly proportioned to $0.256$-second and $0.064$-second bins using
	linear interpolation, in order to correspond to the CTIME data \citep{Meegan}.

\begin{figure*}[t]
\centering
\includegraphics[height=1.4\columnwidth, angle=270]{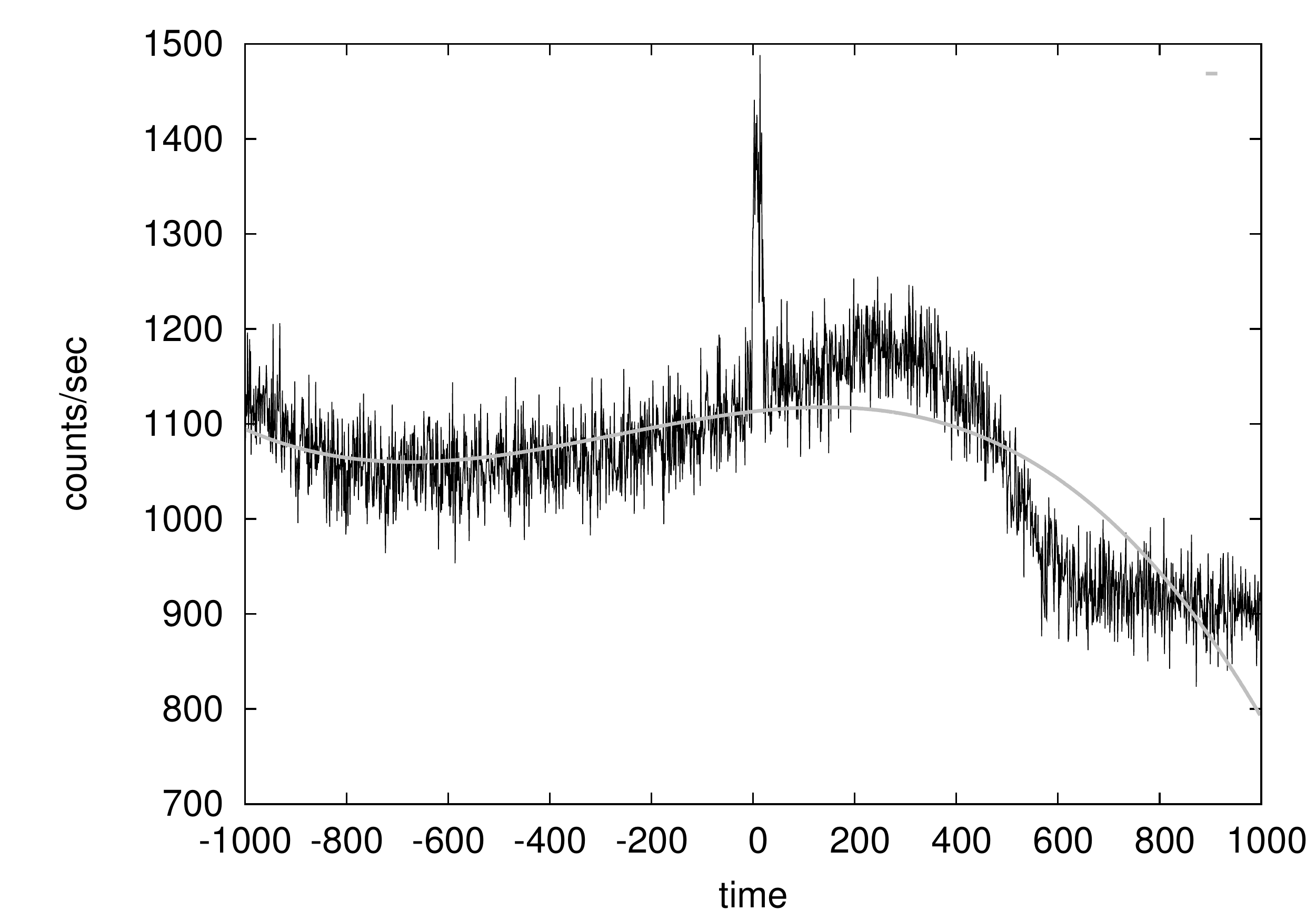}
\caption{Lightcurve of the Fermi burst 091030613 measured by the 3rd
	GBM-detector, without any background filtering, with 1-second bins. The grey
	line is a fitted polynomial function of time of order 3. } \label{fig:lc.eps}
\end{figure*}

	Using GBM data, 1-second bins and summarizing the counts in the
	channels between 11.50--982.23 keV, one can plot a GBM lightcurve as
	shown in Fig.~\ref{fig:lc.eps}.

\section{Direction Dependent Background Fitting (DDBF)}

	Analyzing ancillary spacecraft and other directional data we have
	found the following $(\mathbf{x}_i)$ variables, which seem to
	contribute to the variation of the background: celestial distance
	between burst and detector orientation, celestial distance between Sun
	and detector orientation, rate of the Earth-uncovered sky and time \citep{AA}.  

	We use the method of General Least Square for multidimensional fitting
	of the $\mathbf{y_i}$ counts to the corresponding $(\mathbf{x})$
	explanatory variables.  The maximum likelihood estimate of the model
	parameters $a_k$ is obtained by minimizing the quantity of\\
	\begin{equation}
	\chi^2 = \Sigma_{i=1}^N \left( \frac{y_i- \Sigma_{k=1}^{M}a_k X_k(\mathbf{x}_i) }{\sigma_i} \right)^2
	\label{eq:chi2}
	\end{equation}

	The matrix of $A$ and vector $\mathbf b$ are:
	\begin{equation}
		\displaystyle
		A_{ij} = \frac{X_j(\mathbf{x}_i)}{\sigma_i}, \hspace{20pt} b_{i} = \frac{y_i}{\sigma_i}.
		\label{eq:design}
	\end{equation}

	Minimizing $\chi^2$ leads us to the following equation:
	\begin{equation}
		\mathbf{a} = (\mathbf{A}^T \mathbf{A})^{-1} \mathbf{A}^T \mathbf{b},
		\label{eq:solution}
	\end{equation}
	, where $\mathbf{A}^T$ means the transpose of $\mathbf{A}$, and the
	expression $(\mathbf{A}^T \mathbf{A})^{-1} \mathbf{A}^T$ are called
	\textsl{generalized inverse} or \textsl{pseudoinverse} of $\mathbf{A}$. 
	For calculating the pseudoinverse of the design matrix $\mathbf{A}$ we
	used Singular Value Decomposition (SVD): $\mathbf{A} =
	\mathbf{US}{V}^T$.  Using $\mathbf{U}$ and $\mathbf{V}$, the
	pseudoinverse of $\mathbf{A}$ can be obtained as
        \begin{equation}
	      pinv(\mathbf{A}) = (\mathbf{A}^T \mathbf{A})^{-1} \mathbf{A}^T = \mathbf{VS}^{-1}\mathbf{U}^T.
	      \label{eq:pseudo}
        \end{equation}

	Since we want to have a method for all the Fermi bursts, we define our 
	model to be quite comprehensive.
	Let us have $y(\mathbf{x}_i)$ as the function of
	$\mathbf{x}_i=(x^{(1)}_i,x^{(2)}_i,x^{(3)}_i,x^{(4)}_i)$ of order 3, so
	the basis functions $X_k(\mathbf{x}_i)$ (and columns of the design
	matrix) consist of every possible products of the components
	$x^{(l)}_i$ up to order 3. That means that we have $M=k_{max}=35$ basis
	functions and $a_1$, $a_2$...$a_{35}$ free parameters. It is sure that
	we do not need so many free parameters to describe a simple
	background.  Computing the pseudoinverse we need the reciprocal of the
	singular values in the diagonals of $\mathbf{S}^{-1}$, but if we
	compute the pseudoinverse of $\mathbf{A}$, the reciprocals of the tiny
	and not important singular values will be unreasonably huge and enhance
	the numerical roundoff errors as well. This problem can be solved
	defining a \textsl{limit} value, below which reciprocals of singular
	values are set to zero \citep{AA}. 

	One cornerstone of the fitting algorithm described above is the definition of
	the boundaries which divide the interval of the burst and the intervals of the
	background. In this work, we follow the common method of using user-selected
	time intervals \citep{catalogue}. 
	But unlike \citet{catalogue}, using the position data allows us to fit the total
	CTIME file instead of selecting two small intervals around the burst \citep{AA}.

\section{Model selection}

	The Akaike Information Criterion (AIC) is a commonly used method of
	choosing the right model to the data \citep{AIC}. Assume that we have
	$M$ models so that the $k$th model has $k$ free parameters ($k=1...M$).
	When the deviations of the observed values from the model are normally
	and independently distributed, every model has a value AIC$_k$ so that
        \begin{equation}
	      \mathrm{AIC}_k=N\cdot \log{\frac{RSS_k}{N}} + 2\cdot k
	      \label{eq:AIC}
        \end{equation}
	, where $RSS_k$ is the Residual Sum of Squares from the estimated model
	($RSS=\Sigma_{i=1}^N \left( y_i- y(\mathbf{x}_i,k) \right)^2$), $N$ is
	the sample size and $k$ is the number of free parameters to be
	estimated.  Given any two estimated models, the model with the lower
	value of AIC$_k$ is the one to be preferred. Given many, the one with
	lowest AIC$_k$ will be the best choice: it has as many free parameters
	as it has to have, but not more.

	We loop over the pseudoinverse operation and choose $S_{kk}$ as the limit
	of singular values in the $k$th step,  and compute the corresponding
	AIC$_k$.  The number of singular values which minimalize the AIC$_k$ as
	a function of $k$ will be the best choice when calculating the
	pseudoinverse \citep{AA}.

       As an example we analyze the lightcurve of GRB~090113778.  GRB~090113 is a
       long burst with T$_{90}^{cat}$=17.408$\pm$3.238 s in the GBM Catalogue,
       here we show Detector 0 data:
       \begin{figure}[h!]\begin{center}
	      {\includegraphics[width=.8\columnwidth,angle=270]{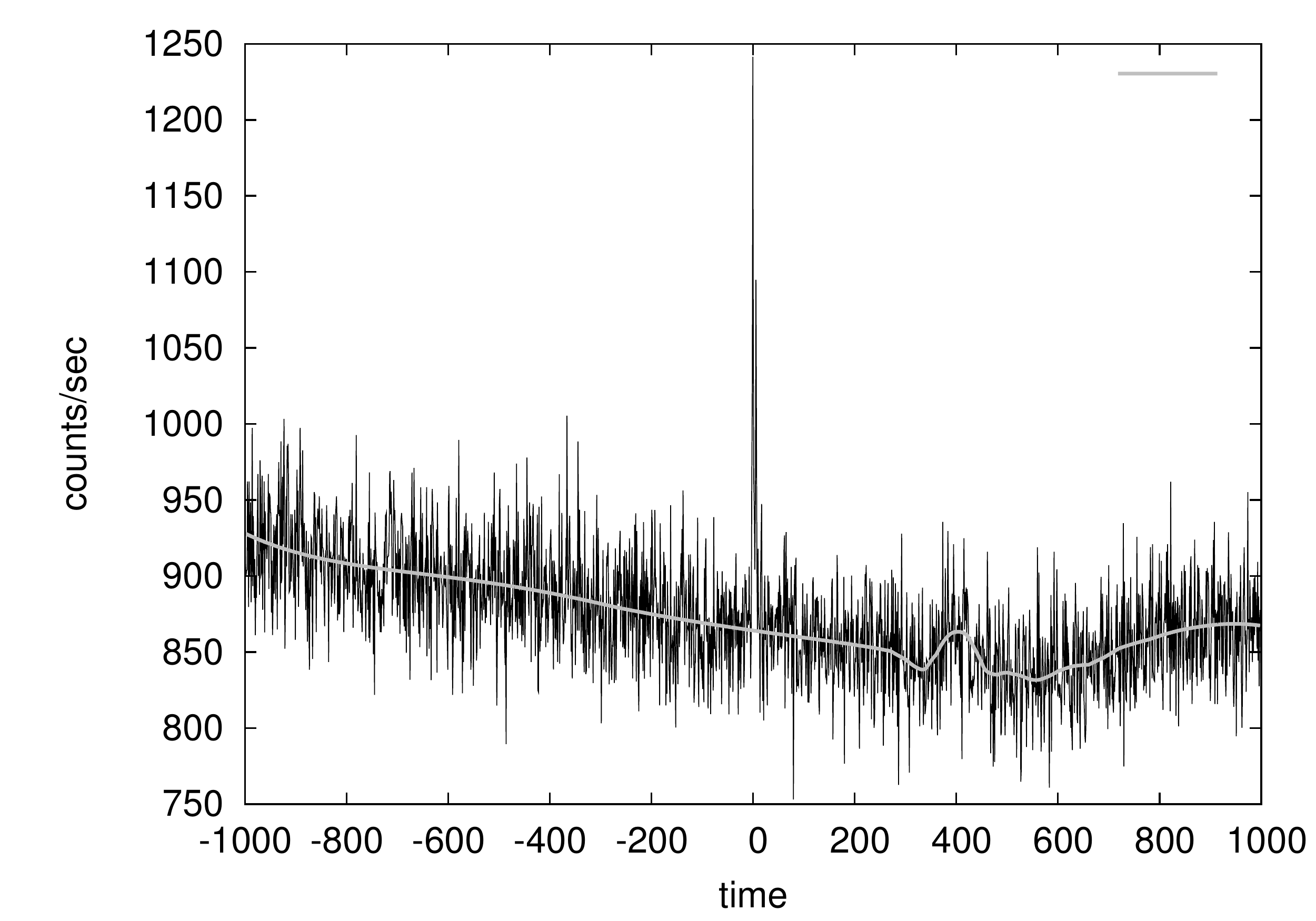}
	      \includegraphics[width=.8\columnwidth,angle=270]{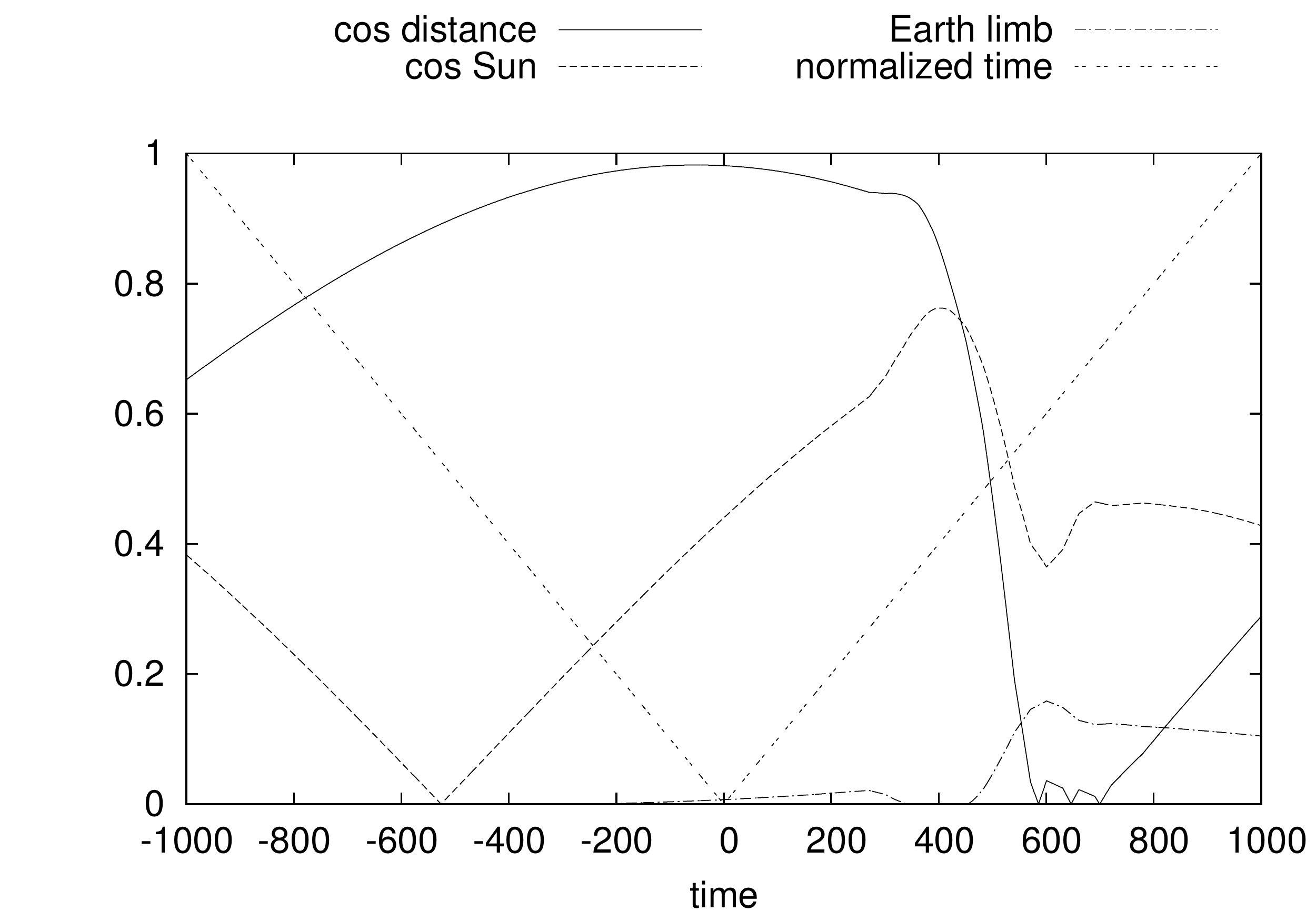}}
		\caption{\small{ 
		\textsl{Up:} Lightcurve of the Fermi GRB~090113778 measured by
		the triggered GBM-detector '8', and the fitted background with
		a grey line. Burst interval (secs): [-20:40].  
		\textsl{Down:} Underlying variables (absolute values).  
		}
		} 
		\label{fig:100130777}
       \end{center}\end{figure}

       \begin{figure}[h!]\begin{center}
	      {\includegraphics[width=.8\columnwidth,angle=270]{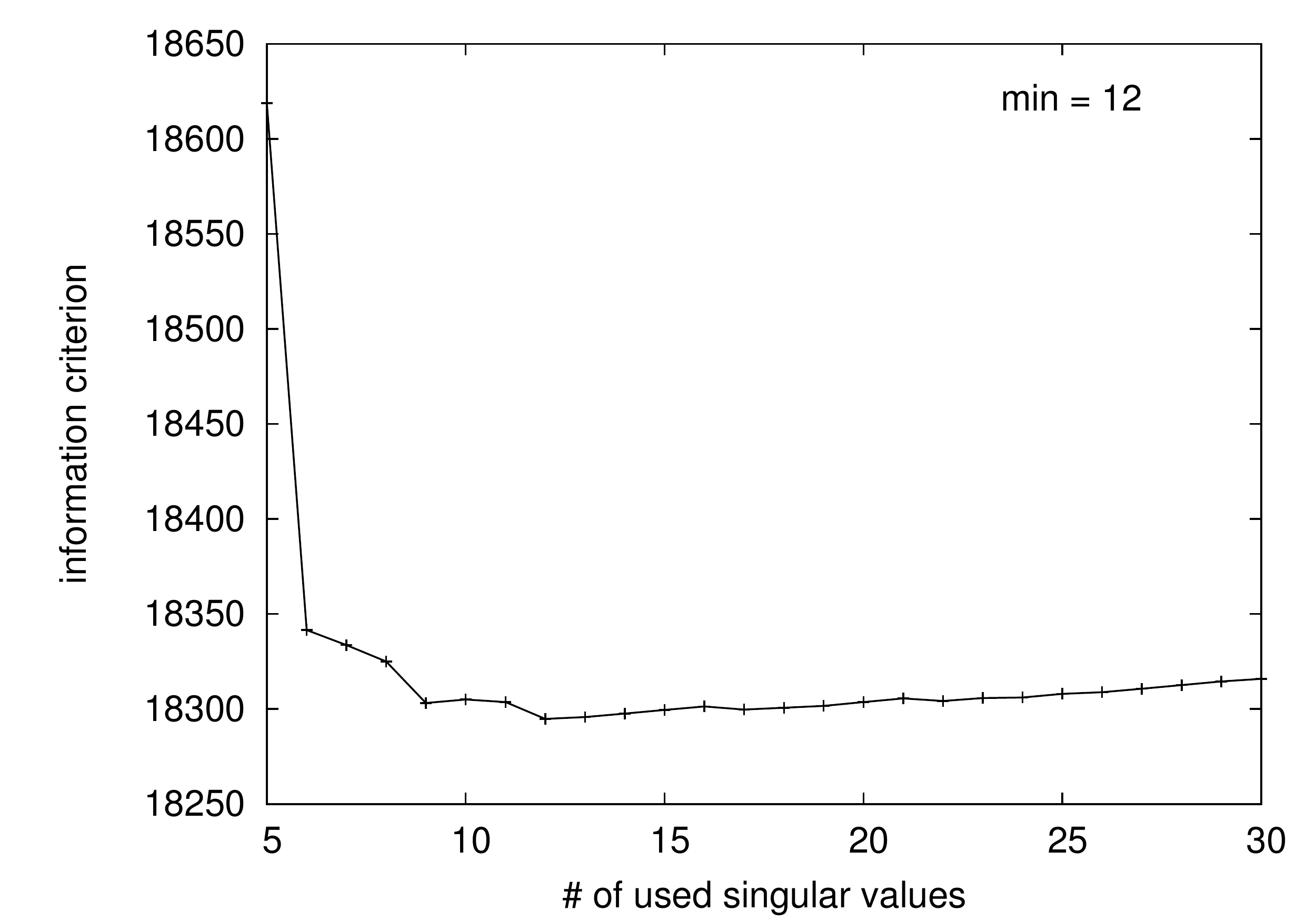}
	      \includegraphics[width=.8\columnwidth,angle=270]{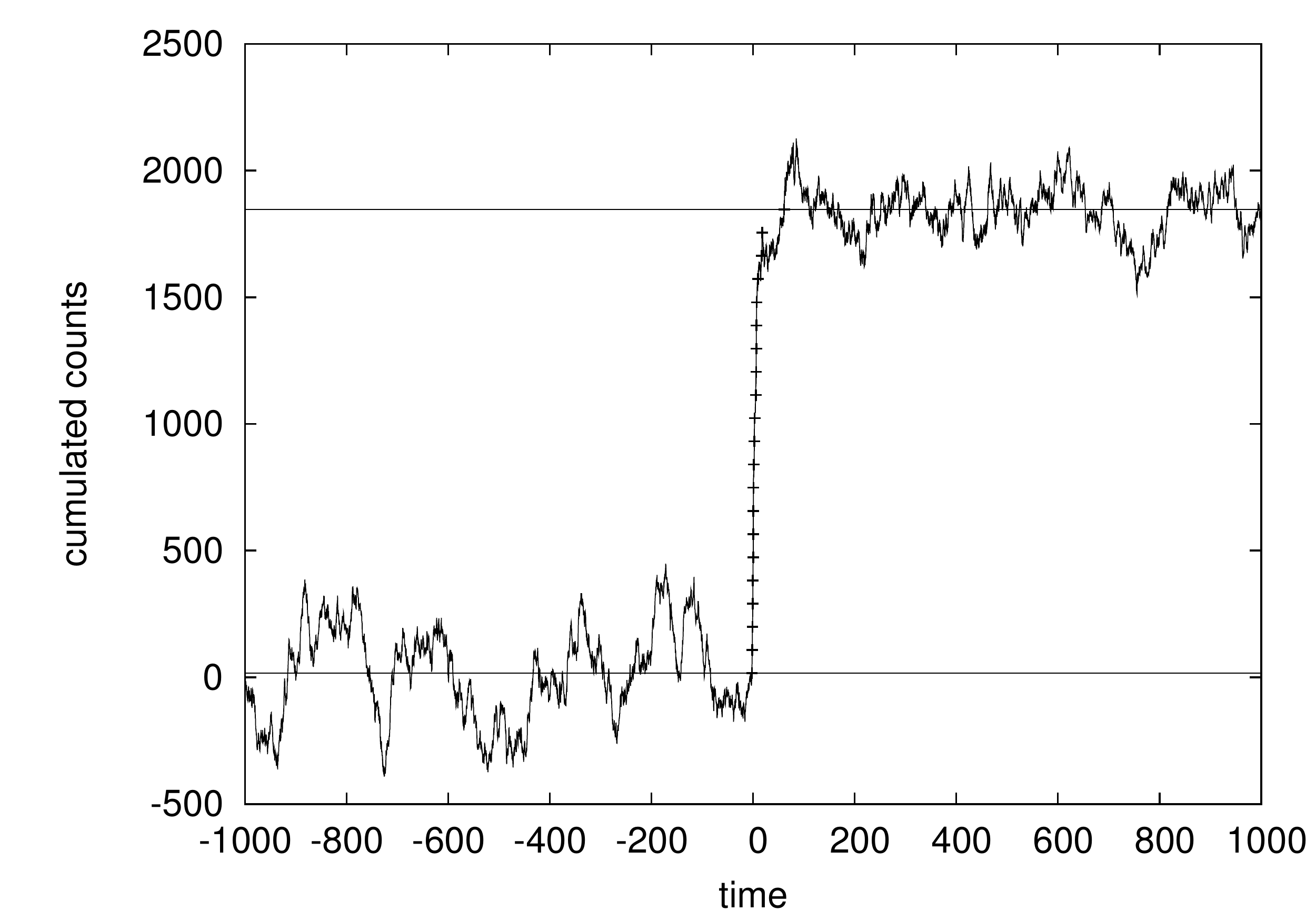}}
				\caption{\small{ 
		\textsl{Down left:} Akaike Information Criterion.  A model with 12 singular values was selected.
		\textsl{Down right:} Cumulated lightcurve of GRB~090113778.
		}}
		\label{fig:100130777int}
       \end{center}\end{figure}
	
       Fig.~\ref{fig:100130777} has some extra counts around 400 and 600
       seconds. Both of them can be explained with the variation of the
       underlying variables, that is, the motion of the satellite. These are
       not GRB signals!

\section{Error analysis}

      The DDBF method is too complicated to give a simple expression
for the error of $T_{90}$ using general rules of error propagation. We
therefore 
repeated the process for 1000 MC simulated data.  

\subsection{GRB~090113778}
\begin{figure}[h!]\begin{center}
  {
  \includegraphics[width=.7\columnwidth]{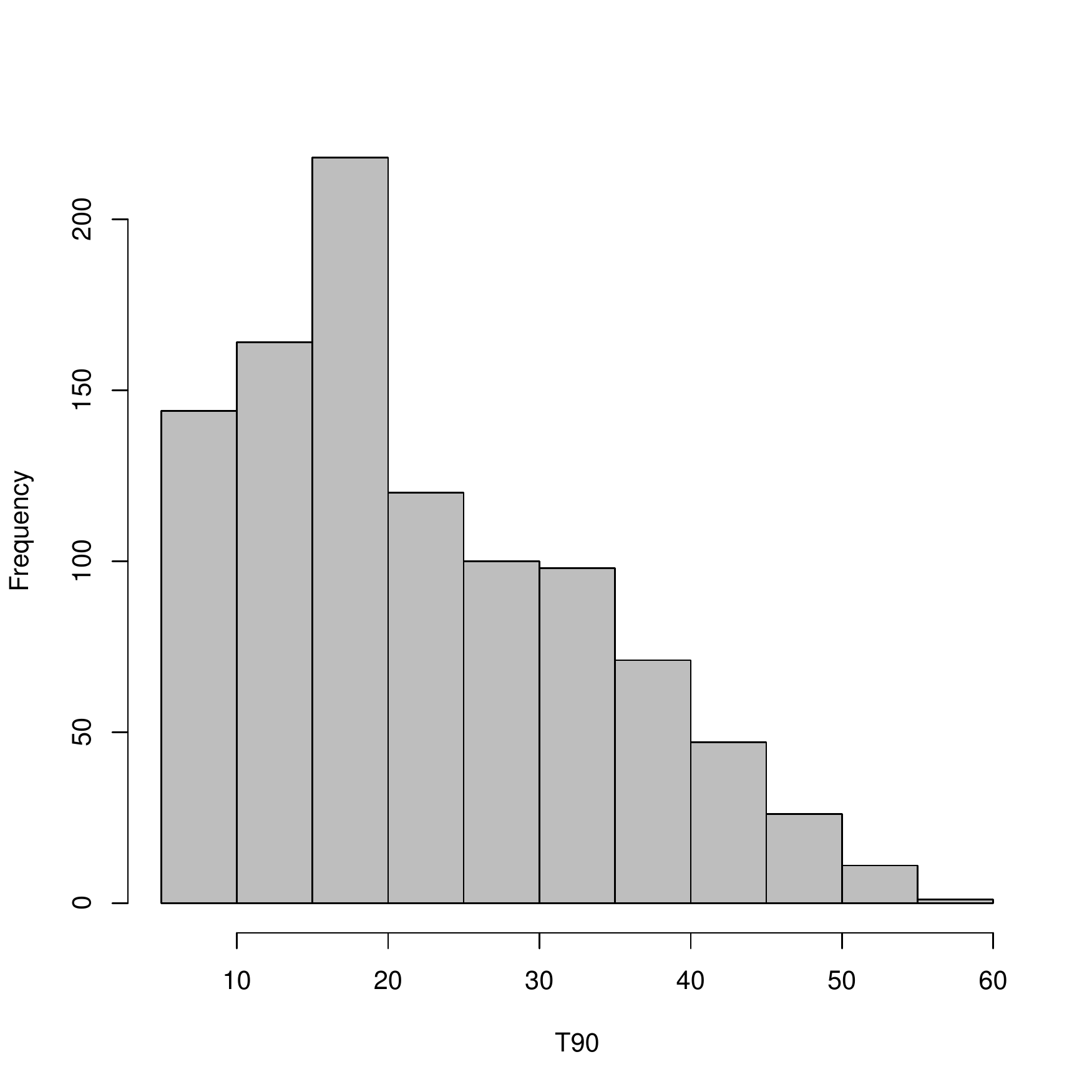}
  \includegraphics[width=.7\columnwidth]{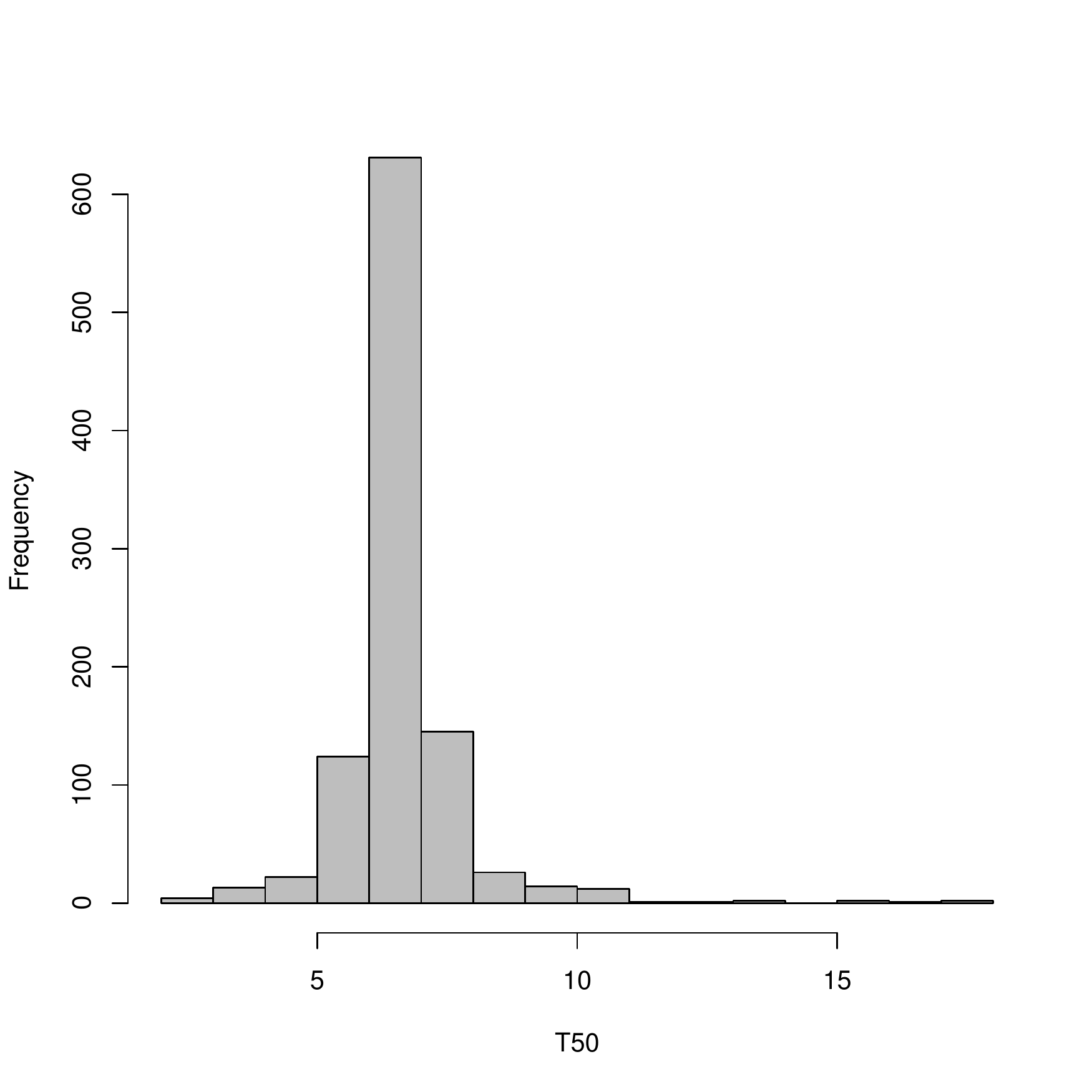}
  }
  \caption{\small{Distibution of the $T_{90}$ (up) and $T_{50}$ (down) values obtained from the MC simulated data for Fermi burst 090113778. }}
  \label{fig:pois90} 
\end{center}\end{figure}

Distribution of the
Poisson-modified $T_{90}$ and $T_{50}$ values are shown in
Fig.~\ref{fig:pois90} for GRB090113778.

Fig.~\ref{fig:pois90} shows the Monte Carlo simulated distributon of the duration values of GRB~090113778. Based on this, we give confidence intervals corresponding to 68\% (1$\sigma$ level). The result is given in Table~\ref{tab:err}. Table~\ref{tab:err}. also contains some of our other, preliminary results using DDBF. For more details, see our forthcoming paper \citet{AA}.

\subsection{GRB~091030613}
\begin{figure}[h!]\begin{center}
  {
  \includegraphics[width=.43\columnwidth]{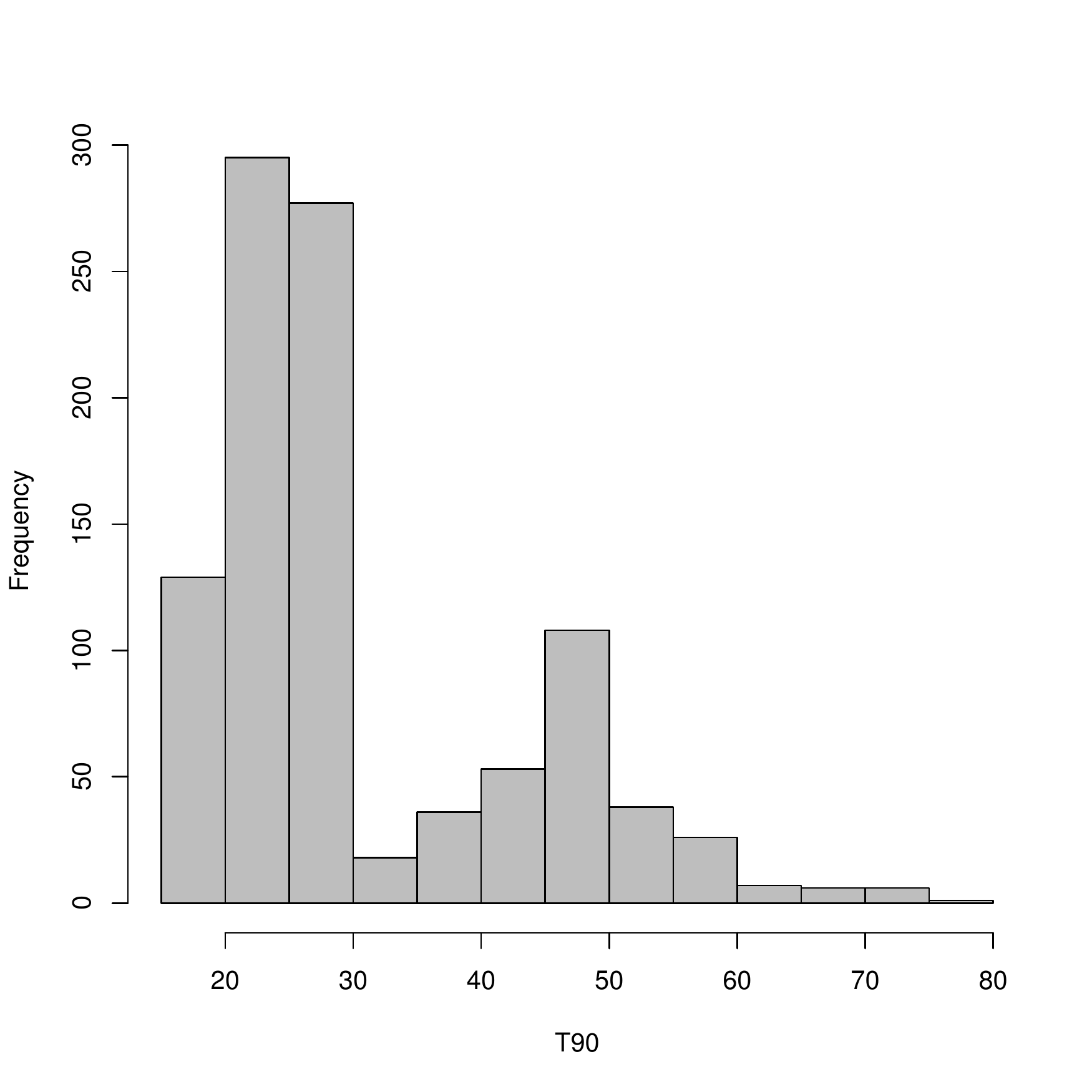}
  \includegraphics[width=.43\columnwidth]{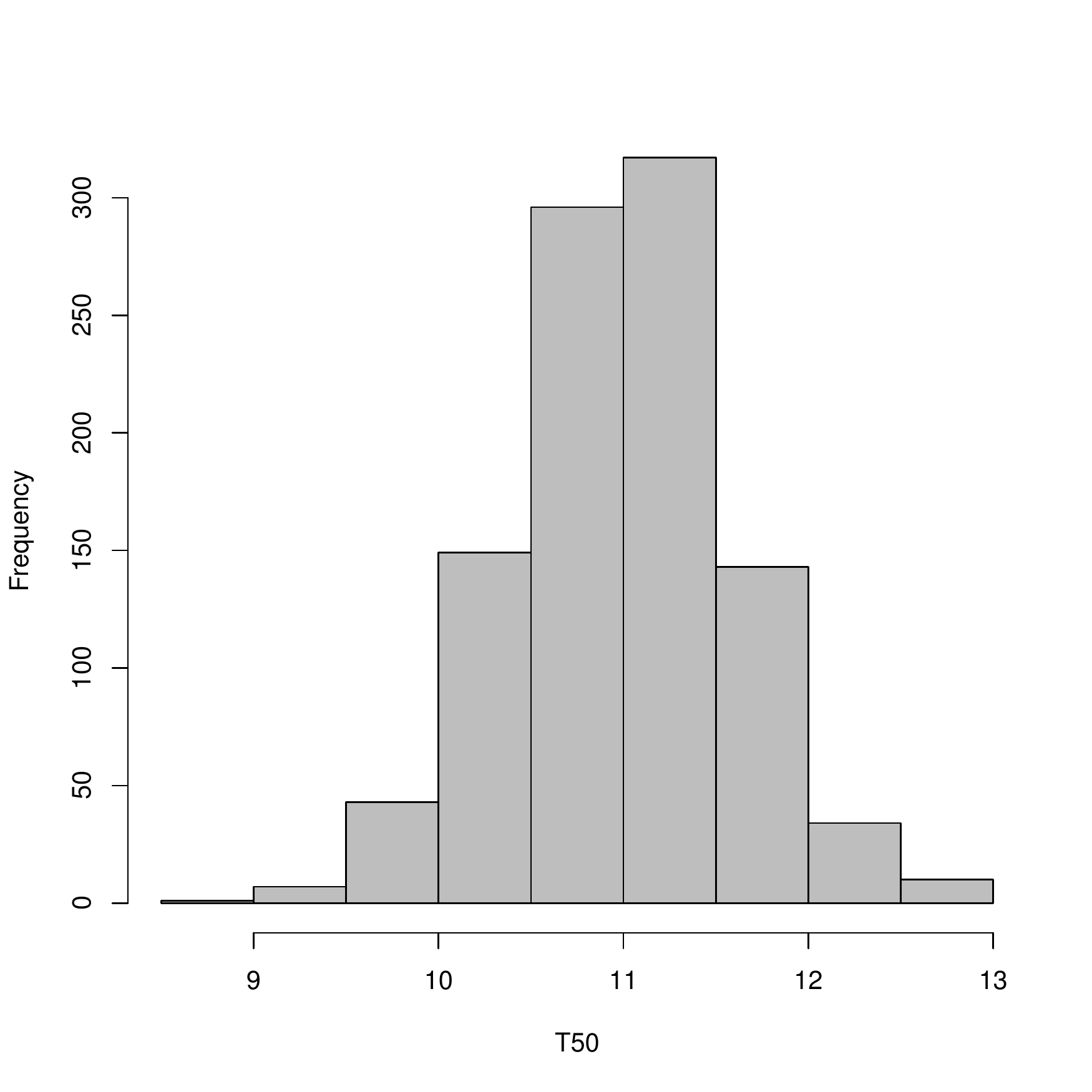}
  }
  \caption{\small{Distibution of the $T_{90}$ (left) and $T_{50}$ (right) values obtained from the MC simulated data for Fermi burst 091030613.  }}
  \label{fig:pois091030613} 
\end{center}\end{figure}

Distribution of the
Poisson-modified $T_{90}$ and $T_{50}$ values are shown in
Fig.~\ref{fig:pois091030613} for GRB091030613.

Fig.~\ref{fig:pois90} shows two significant peaks around 22 and 47 seconds.
The first peak at 22 seconds corresponds to the measured $T_{90}$ value.
However, in some cases of the Poison noise simulation, the measured $T_{90}$
value is systematically longer: that is because this burst has a little pulse
around 47 seconds (see Fig.~\ref{fig:lc.eps}.).
There is no sign of this second peak in the $T_{50}$ distribution, as it 
is more robust. 

\subsection{Preliminary results}

\begin{figure}[h!]\begin{center}
  {  \includegraphics[width=.43\columnwidth]{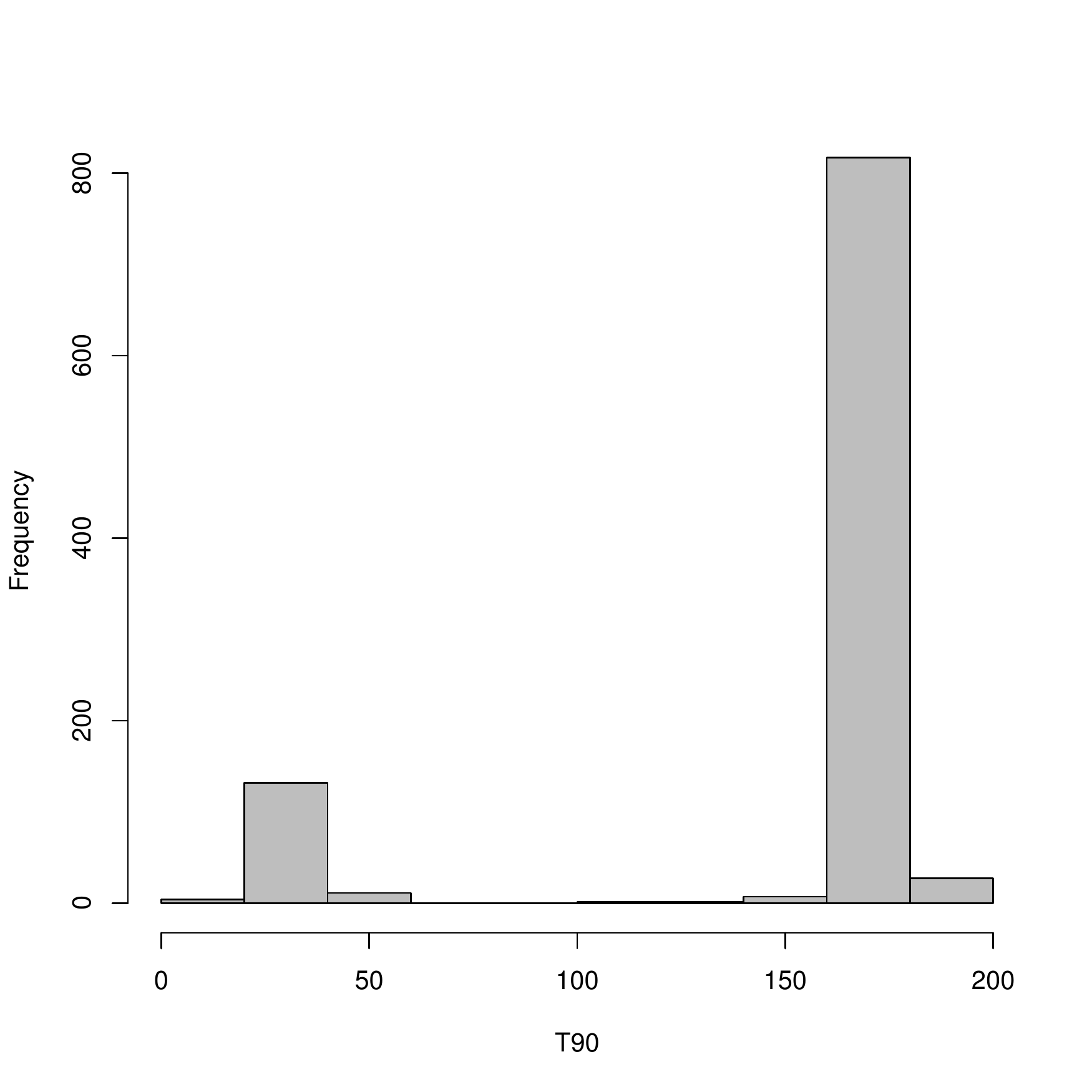}
  \includegraphics[width=.43\columnwidth]{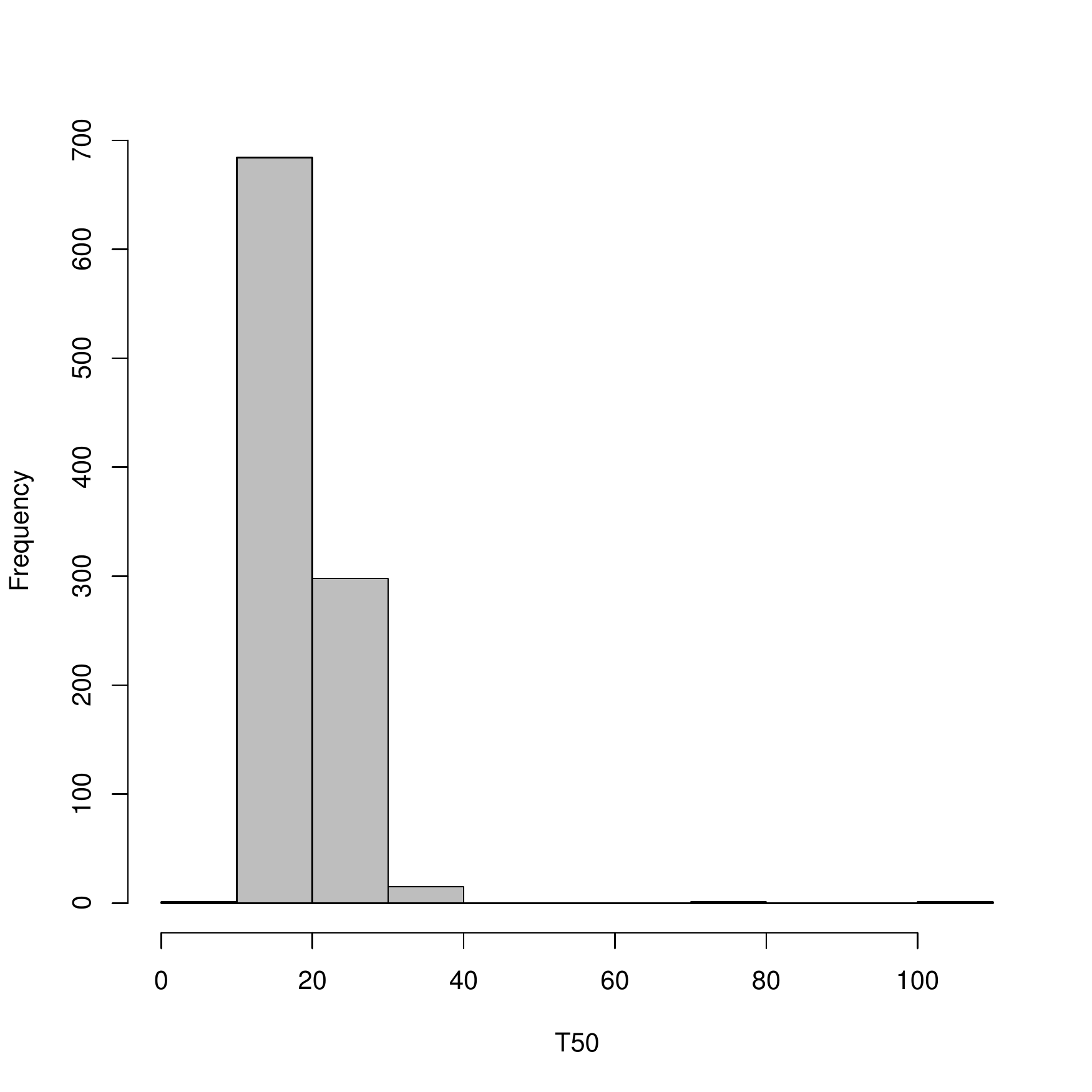}}
  \caption{\small{Distibution of the $T_{90}$ (left) and $T_{50}$ (right) values obtained from the MC simulated data for Fermi burst 081009360, GBM detector '8'.  }}
  \label{fig:pois081009360} 
\end{center}\end{figure}

\begin{figure}[h!]\begin{center}
  {  \includegraphics[width=.43\columnwidth]{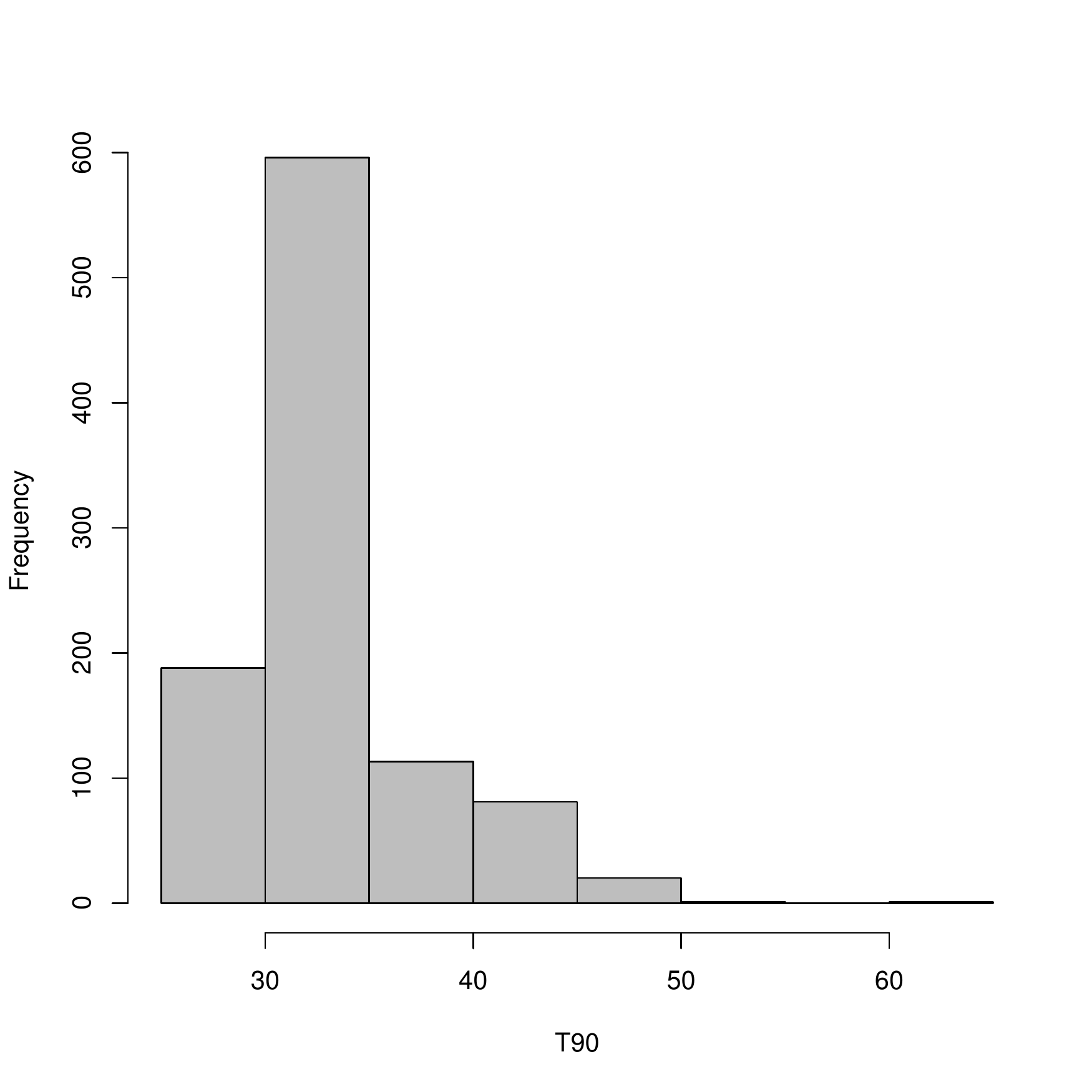}
  \includegraphics[width=.43\columnwidth]{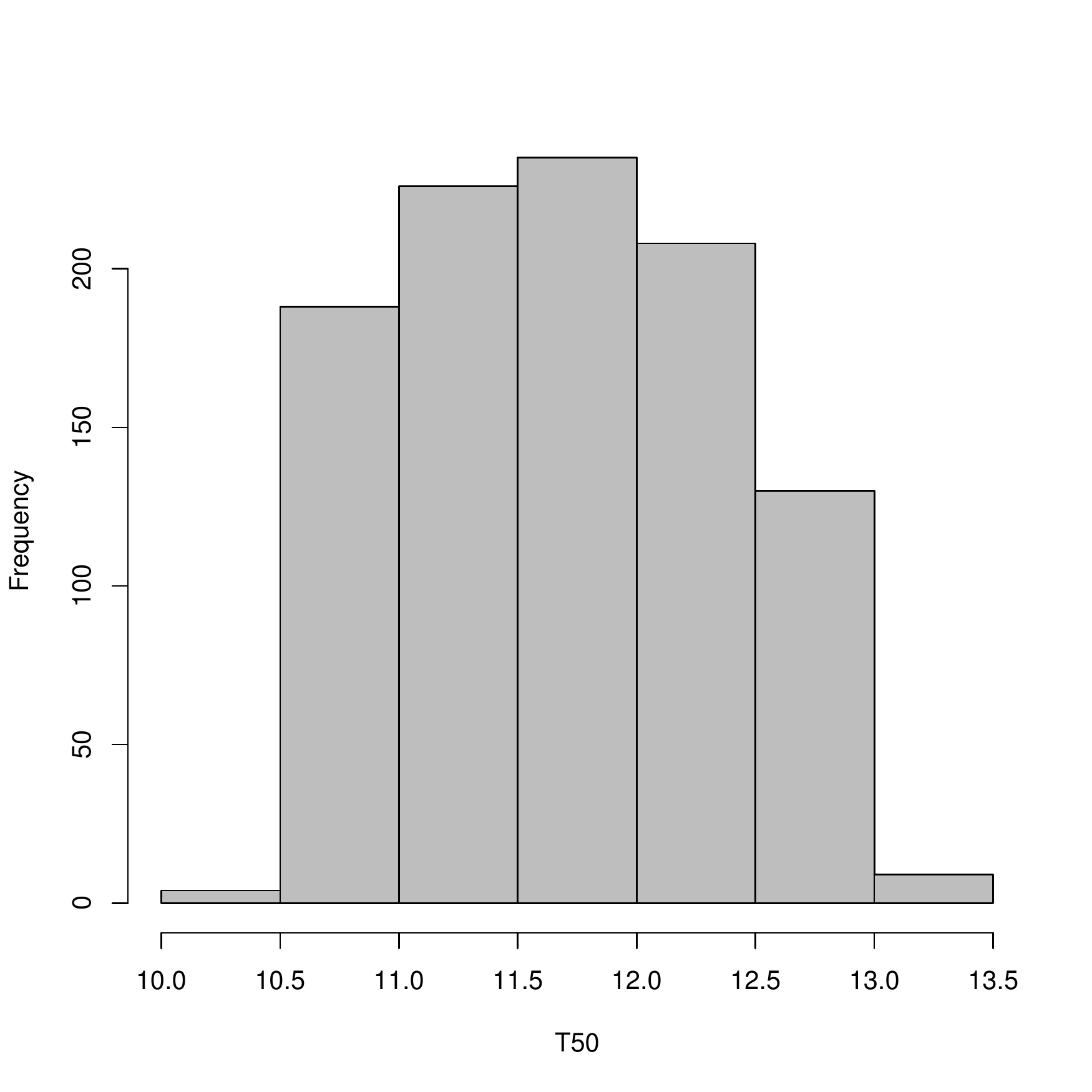}}
  \caption{\small{Distibution of the $T_{90}$ (left) and $T_{50}$ (right) values obtained from the MC simulated data for Fermi burst GRB 090102122, GBM detector 'a'.  }}
  \label{fig:poisT90090102122na} 
\end{center}\end{figure}

\begin{figure}[h!]\begin{center}
  {  \includegraphics[width=.43\columnwidth]{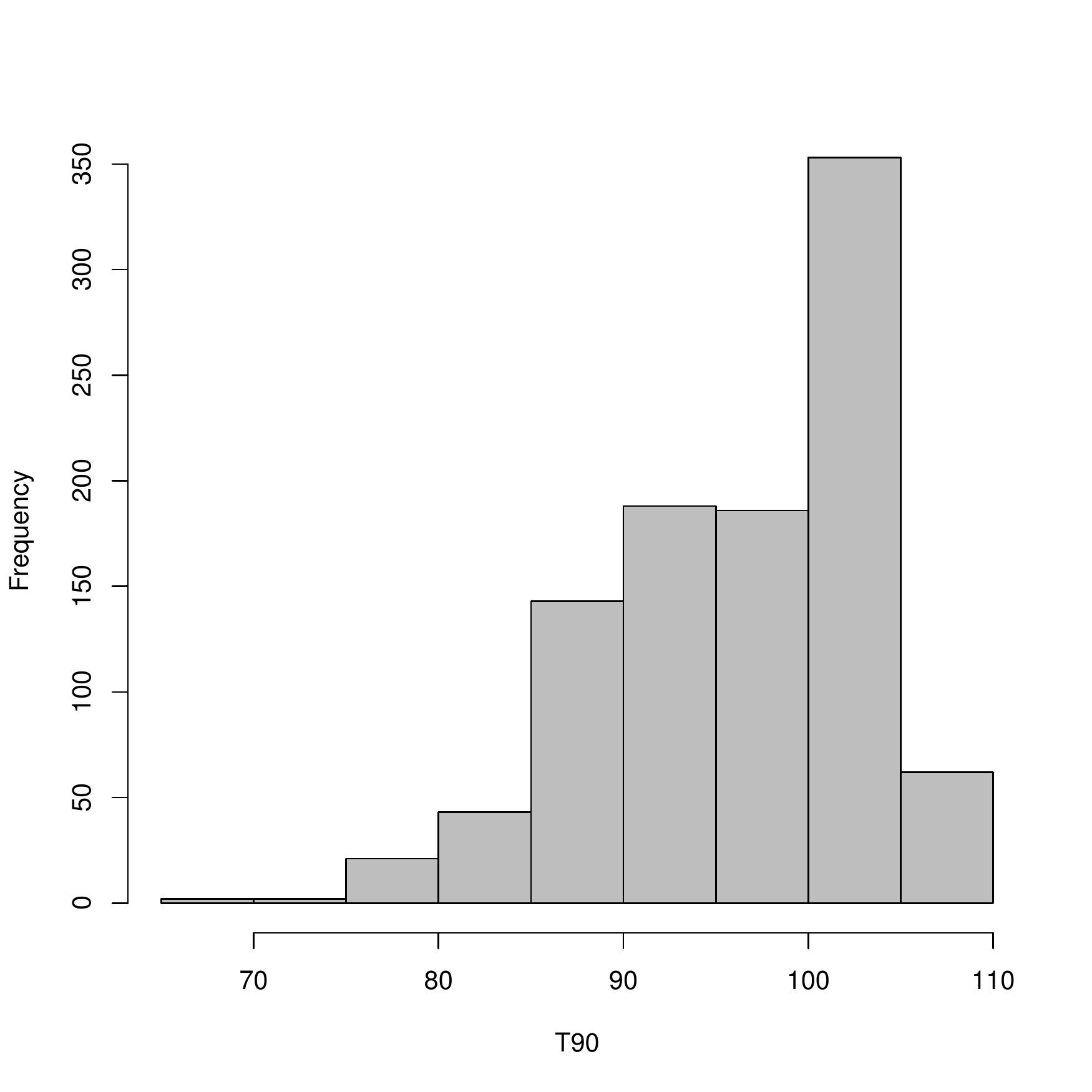}
  \includegraphics[width=.43\columnwidth]{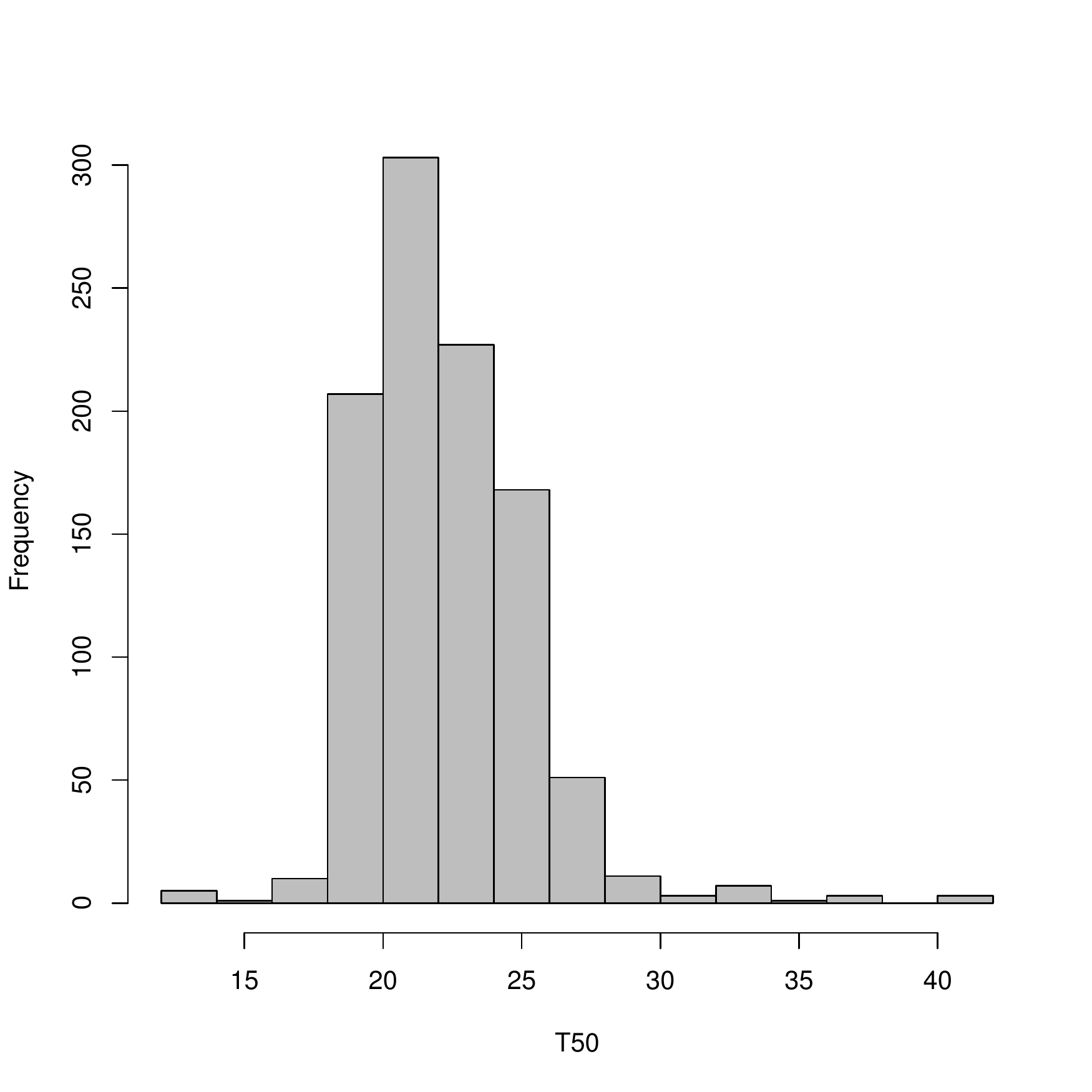}}
  \caption{\small{Distibution of the $T_{90}$ (left) and $T_{50}$ (right) values obtained from the MC simulated data for Fermi burst 090618353, GBM detector '7'.  }}
  \label{fig:poisT90090618353n7} 
\end{center}\end{figure}

\begin{figure}[h!]\begin{center}
  {  \includegraphics[width=.43\columnwidth]{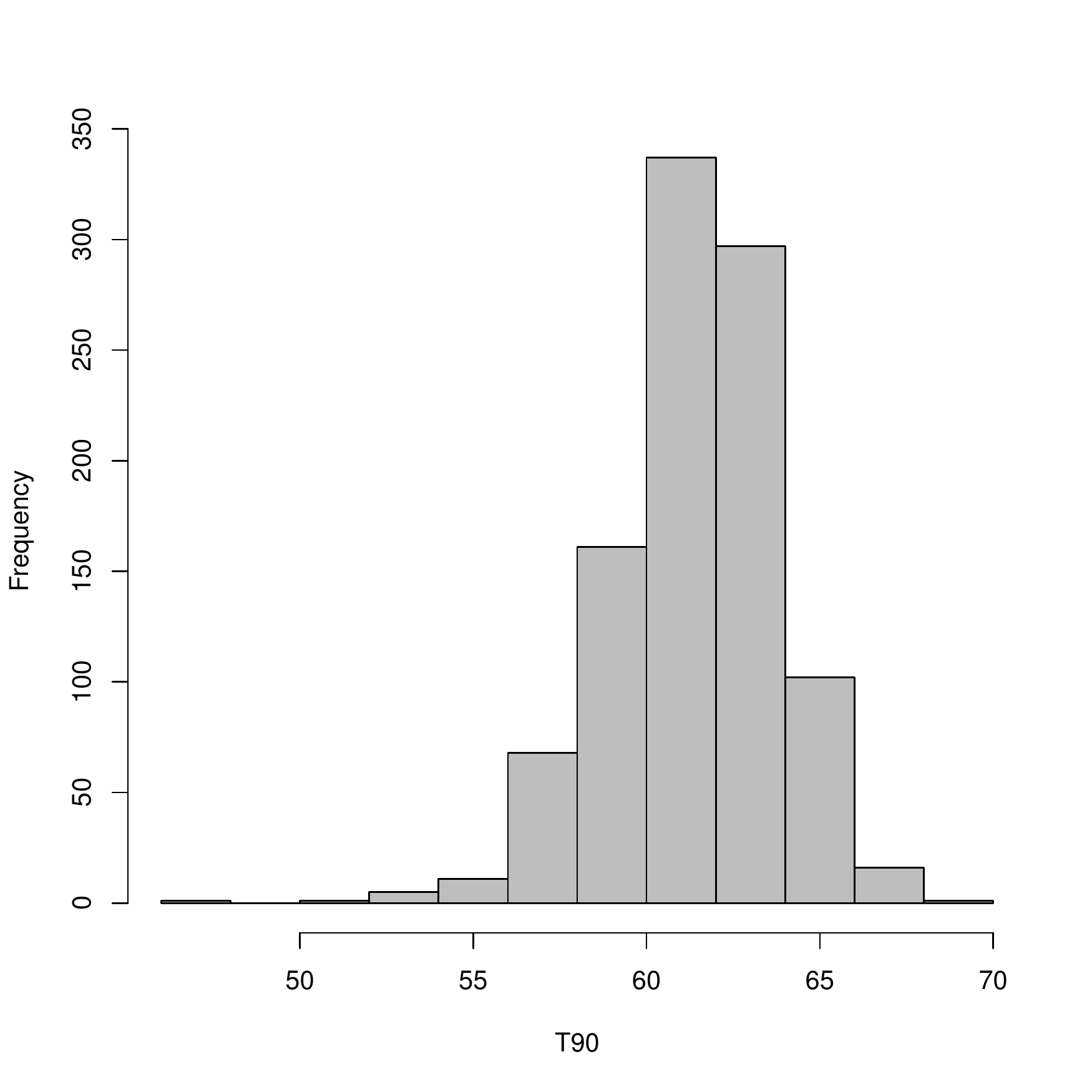}
  \includegraphics[width=.43\columnwidth]{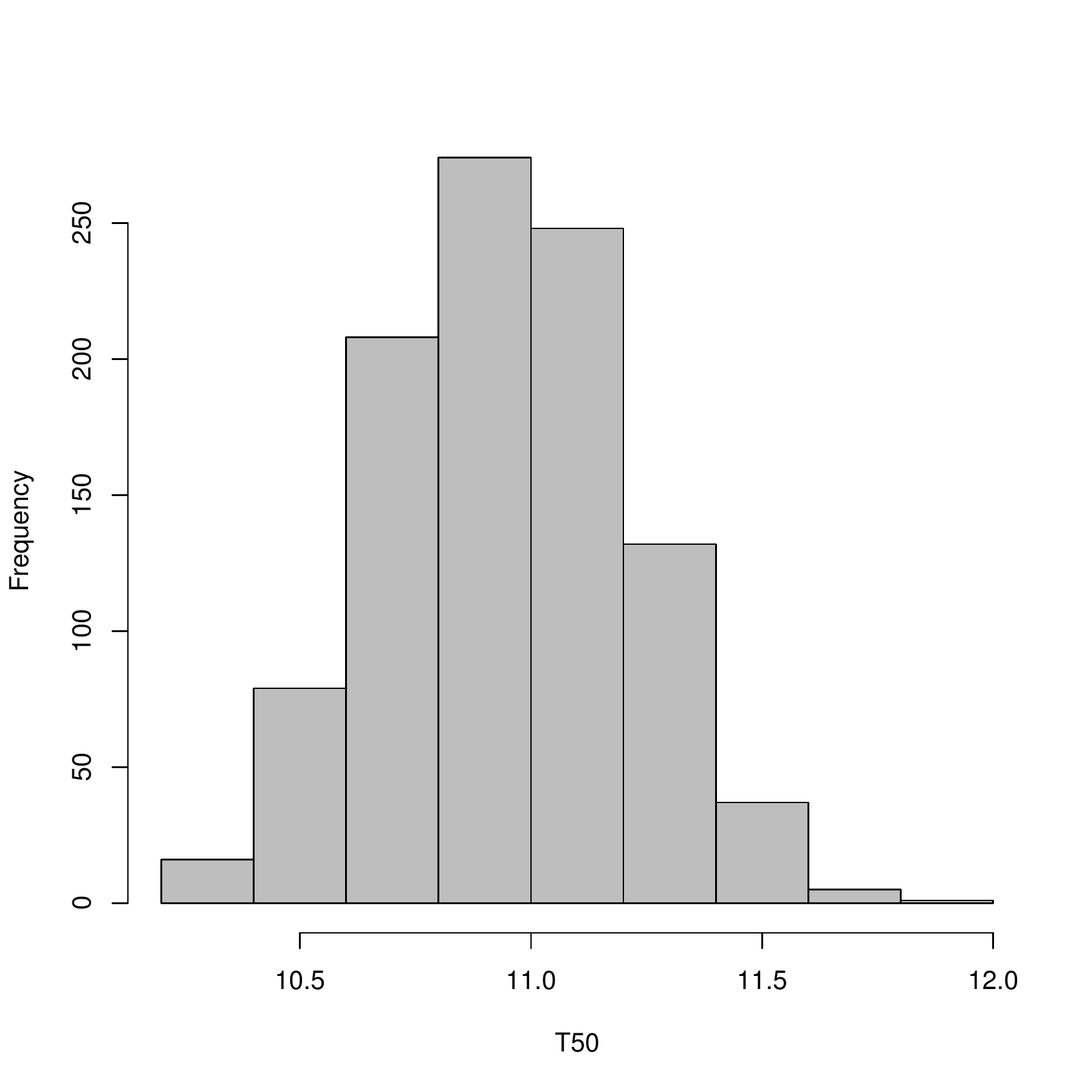}}
  \caption{\small{Distibution of the $T_{90}$ (left) and $T_{50}$ (right) values obtained from the MC simulated data for Fermi burst 090828099, GBM detector '5'.  }}
  \label{fig:poisT90090828099n5} 
\end{center}\end{figure}

\begin{figure}[h!]\begin{center}
  {  \includegraphics[width=.43\columnwidth]{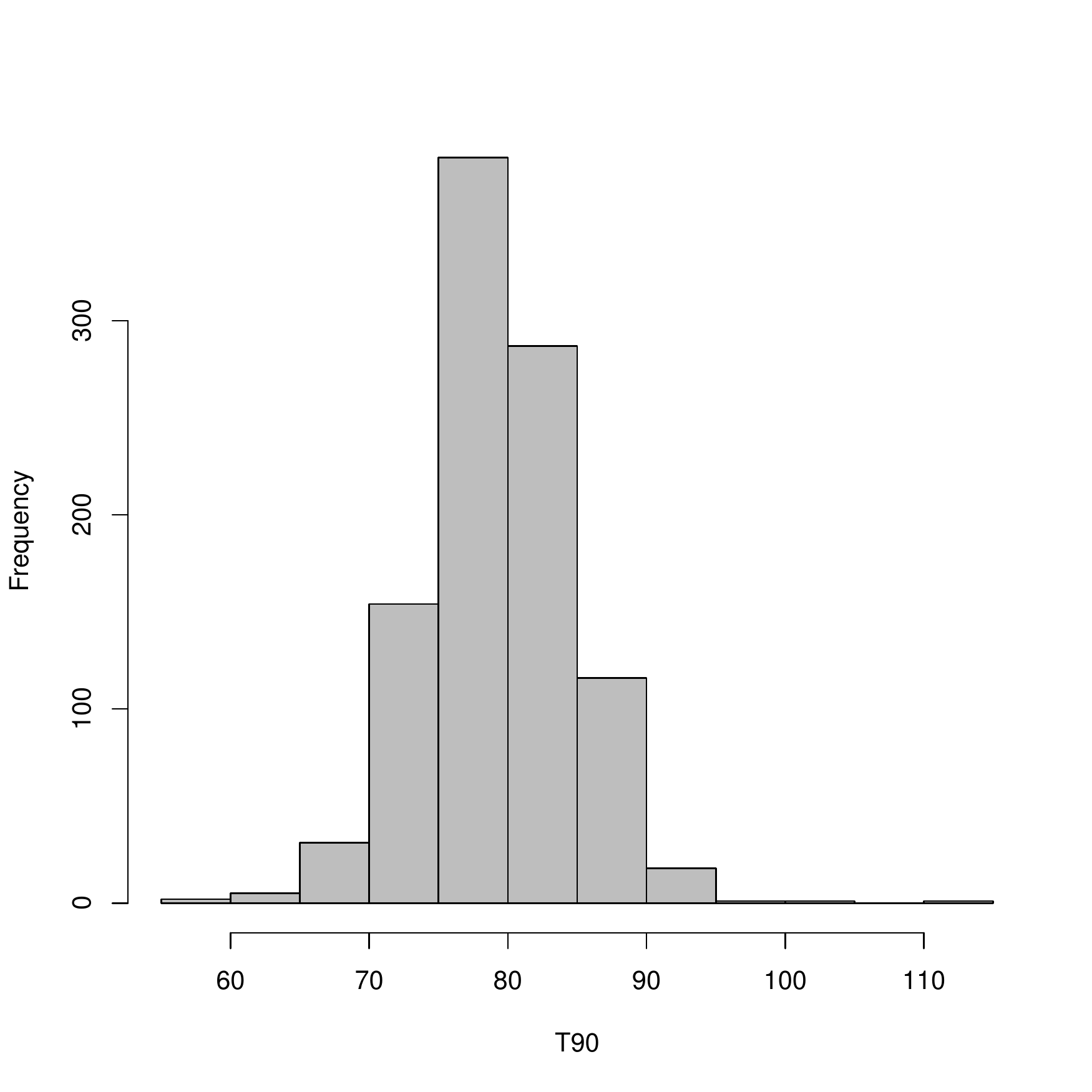}
  \includegraphics[width=.43\columnwidth]{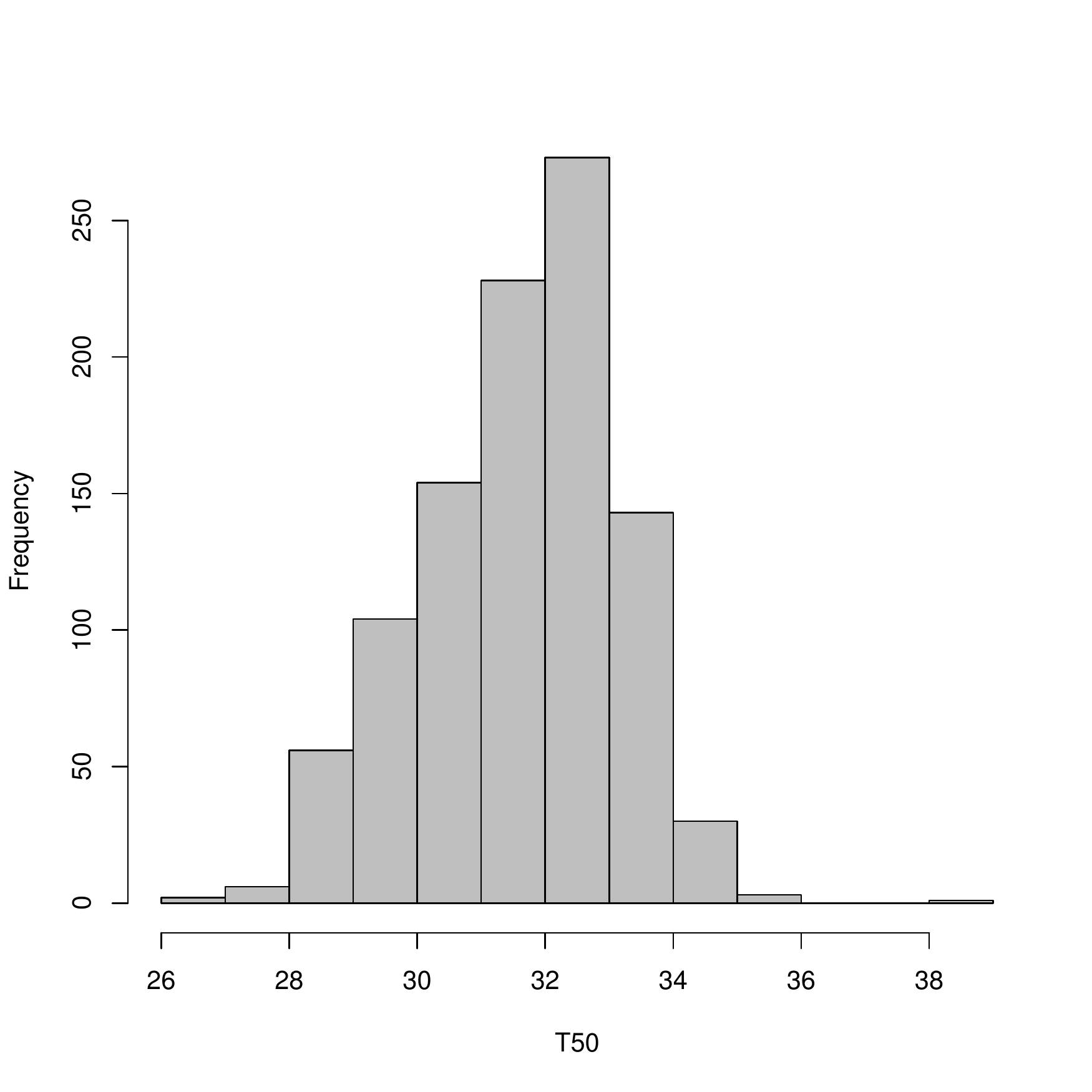}}
  \caption{\small{Distibution of the $T_{90}$ (left) and $T_{50}$ (right) values obtained from the MC simulated data for Fermi burst 100130777, GBM detector 'b'.  }}
  \label{fig:poisT90100130777nb} 
\end{center}\end{figure}

In Figures~\ref{fig:pois081009360}-\ref{fig:poisT90100130777nb} show the MC simulated $T_{90}$ and $T_{50}$ values for the GRBs given in Table~\ref{tab:err}.

	\begin{center}
\begin{table*}
{
\begin{tabular}{|l||c|c|c|c||c|c|c|c|}\hline 
	Burst & $T_{90} (s) $ & \multicolumn{2}{c|}{errors $(s)$} & $T_{90}^{catalog} (s) $ & $T_{50} (s) $ & \multicolumn{2}{c|}{errors $(s)$} & $T_{50}^{catalog} (s)$ \\ \hline \hline
	081009690 & 176.228 & ${+1.357}$&${-9.477}$ & 176.191 & 15.852 & $ {+3.006}$&${-2.350}$ & 25.088 \\ \hline
	090102122 & 29.756 & $ {+2.971}$&${-1.198}$ & 26.624 & 10.859 & $ {+0.531}$&${-0.556}$ & 9.728 \\ \hline
	090113778 & 19.679 & $ {+10.883}$&${-6.421}$ & 17.408 & 6.408 & $ {+0.498}$&${-0.344}$ & 6.141 \\ \hline
	090618353 & 103.338 & $ {+3.842}$&${-6.725}$ & 112.386 & 22.827 & $ {+2.201}$&${-1.530}$ & 23.808 \\ \hline
	090828099 & 63.608 & $ {+1.467}$&${-1.652}$ & 68.417 & 11.100 & $ {+0.198}$&${-0.194}$ & 10.752 \\ \hline
	091030613 & 22.609 & $ {+13.518}$&${-4.522}$ & 19.200 & 10.770 & $ {+0.388}$&${-0.424}$ & 9.472 \\ \hline
	100130777 & 80.031 & $ {+3.755}$&${-3.485}$ & 86.018 & 32.340 & $ {+0.931}$&${-1.363}$ & 34.049 \\ \hline
\end{tabular}}\ \\\  \\
\caption{{Preliminary results and confidence intervals. $T_{90}^{catalog}$ and $T_{50}^{catalog}$ are from the GBM Catalogue \citep{catalogue}. }}
\label{tab:err}
\end{table*}
\end{center}

\section{Summary}

	We summarized the Direction Dependent Background Fitting (DDBF) algorithm which
	is designed to filter the background of the Fermi lightcurves on a longer timescale
	(2000 seconds of the CTIME datafile).
	This technique is based on the motion and orientation of the satellite.
	The DDBF considers the position of the burst, the Sun and the Earth
	as well. Based on these position information, we computed three physically
	meaningful underlying variables, and fitted a 4 dimensional hypersurface on
	the background. Singular value decomposition and Akaike information criterion
	was used to reduce the number of free parameters. 

	One of the main advantage of the DDBF method is that it considers only variables with
	physical meaning. Furthermore, this method can fit the total 2000 sec
	CTIME data as opposed to the currently used methods. This features are
	necessary when analyzing long GRBs and precursors, where motion effects
	influence the background rate sometimes in a very extreme way.
	Therefore, not only Sky Survey, but ARR mode GRB's can be analyzed, and
	a possible long emission can be detected.  
	
	A detailed description of the DDBF method, the first resulst, the comparison with the currently used methods, and analysis of the ARR cases will be published soon \citep{AA}.

\bigskip 
\begin{acknowledgments}
This work was supported by OTKA grant K077795, by OTKA/NKTH A08-77719 and A08-77815 grants (Z.B.).
  The authors are grateful to \'Aron Szab\'o, P\'eter Veres for the valuable discussions.
\end{acknowledgments}

\bigskip 

\end{document}